\documentclass[12pt]{article}

\begin{document}

\thispagestyle{empty}
\renewcommand{\thefootnote}{\fnsymbol{footnote}}

\begin{flushright}
{\small
SLAC--PUB--10134\\
October 2003\\}
\end{flushright}

\vspace{.8cm}

\begin{center}
{\bf\Large   
Accelerator Experiments for Astrophysics
\footnote{
This work was supported by the Department of Energy under contract
DE-AC03-76SF00515.}}

\vspace{1cm}
 { \renewcommand{\baselinestretch}{1.0} \Large \normalsize
   Johnny S.T. Ng}

\vspace{10pt}
 { \renewcommand{\baselinestretch}{1.0} \Large \normalsize
{\it Stanford Linear Accelerator Center\\
Accelerator Research Department A, MS 26 \\
P.O. Box 20450, Stanford, CA. 94309-2010, USA.\\
E-mail: jng@SLAC.Stanford.edu}
}

\end{center}
\vfill

\begin{center}
{\bf\large   
Abstract }
\end{center}

\begin{quote}

Many recent discoveries in
astrophysics involve phenomena that are highly complex.
Carefully designed experiments, together with sophisticated computer 
simulations, are required to gain insights into the underlying physics.  We
show that particle accelerators are unique tools
in this area of research, by providing precision calibration data and 
by creating extreme experimental conditions relevant for astrophysics.
In this paper we discuss laboratory experiments that can be carried
out at the Stanford Linear Accelerator Center and implications for
astrophysics.  
\end{quote}

\vfill

\begin{center} 
{\it Invited talk presented at the Joint 28$^{th}$ ICFA Advanced Beam
Dynamics \& Advanced \& Novel Accelerators Workshop on QUANTUM ASPECTS OF
BEAM PHYSICS.}\\
{\it Hiroshima University, Higashi Hiroshima, Japan.}\\
{\it January 7-11, 2003}\\

\end{center}

\newpage



%
\pagestyle{plain}

\section{Introduction}
Recent advances in high-energy astrophysics involve observations of
extremely complex phenomena such as jets
from active galactic nuclei (AGN), gamma-ray bursts (GRB), 
and ultrahigh-energy cosmic rays (UHECR).  Observations of AGN jets, 
consisting of a highly 
collimated stream of material, show that the outflow expands at relativistic 
velocity and spans a distance scale of thousands of light-years.  
The collimation and 
production mechanism, most likely involving the
dynamics of accretion disks around a black hole in the center of the AGN,
are subjects of current research.
Gamma-ray bursts, on the other hand, 
are some of the brightest observed light sources in the universe.
The amount of electromagnetic energy output in a burst is 
equivalent to several times the solar mass released in a matter of seconds.  
The out-flowing materials of a GRB expand at relativistic velocity 
as well, and are possibly collimated, similar to an AGN jet.
The nature of the progenitor and
explanations for their observed characteristics are currently under
debate.  Much of our current understanding of these objects are 
inferred from the properties of the observed radiation.

A strong magnetic field is believed to exist in both the GRB and the 
AGN jets.  The interaction of the relativisticly expanding material with the
environment can lead to nonlinear plasma phenomena that result in the
acceleration of particles to high energies.  
Ultrahigh-energy cosmic rays, with energies observed up to around $10^{20}$ eV,
are believed to come from extra-galactic sources.  
The nature and origin of these cosmic rays 
as well as their acceleration mechanism are still a mystery.

The study of these extreme phenomena requires tremendous effort.  So far,
progress in our understanding has required a combination of observation,
numerical simulations, and theoretical modeling.\cite{sciencemag-rev}
However, astrophysical observations
must be carefully checked for instrumentation effects.
And the complex numerical and theoretical calculations used to 
interpret these observations must be validated.  Thus, it is important
to calibrate the techniques used in the observations and to 
benchmark computer model calculations.  Furthermore,
since observational astrophysics deals with uncontrolled environments,
laboratory experiments able to model the relevant 
extreme conditions would provide
unique insight into the underlying physical mechanisms.

Laboratory studies, ranging from work on atomic spectroscopy, and the
studies of hydrodynamics,
radiation flow, and the equation-of-state using intense 
lasers\cite{remington-takabe},
have been instrumental in astrophysics research.
Recently, it has been suggested that accelerators can be used in the
laboratory investigation of extreme
astrophysical phenomena.\cite{chen-labastro}  
In this paper we discuss possible experiments using intense
particle and photon beams to verify astrophysical observations and to
study relativistic plasma dynamics and
ultrahigh-energy cosmic acceleration mechanisms.
An overview of the accelerator facility at the Stanford Linear Accelerator
Center (SLAC) is given in Section~\ref{sec-0}.
In Section~\ref{sec-1}, we discuss calibration experiments,
focusing on the
discrepancy in the UHECR spectrum measured by two large-aperture cosmic
ray experiments, and describe an experiment that may help resolve it.
In Section~\ref{sec-2}, we discuss laboratory experiments 
that may improve our understanding of the underlying dynamics of 
high-energy astrophysics phenomena.
We conclude with an outlook in Section~\ref{sec-3}.

\section{An Overview of the SLAC Facility}
\label{sec-0}

\begin{figure}[t]
\centerline{\epsfxsize=4.5in\epsfbox{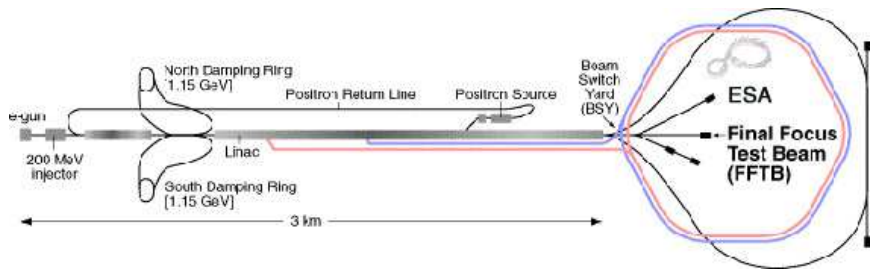}}   
\caption{Layout of the SLAC facility. \label{slac}}
\end{figure}

The 3-km long linear accelerator is the backbone for SLAC's high-energy
physics research program.   
It is capable of delivering electrons and positrons
with 50~GeV energy and 120~Hz repetition rate at $10^{10}$ 
particles per pulse.  Currently it serves
as the injector for the PEP-II storage ring to produce copious amount of
B-meson particles for CP-violation measurements.  It can also deliver beams
to the fix-target experimental hall End-Station A (ESA) and the Final Focus
Test Beam (FFTB).  A schematic layout of the SLAC facility is
shown in Figure~\ref{slac}. 
High intensity photon beams, tunable from X-ray to gamma-ray, can
be derived using a variety of methods, such as undulators, laser-Compton
back-scattering, and bremsstrahlung.  Depending on the required wavelength,
typical fluences of $10^9$ photons per pulse can be provided.

\input epsf.tex

\section{Calibration Experiments}
\label{sec-1}

\subsection{Detector Calibration}
High energy beams from the linac can be used to generate
a variety of secondary and tertiary beams for calibration purposes.
A secondary pion and positron beam with well-defined momentum can be 
generated using a combination
of target and selection magnet system, with a beam intensity that can be
set to below 1 particle per pulse.  With the addition of a tagger magnet, this
secondary beam can be converted into a photon beam with known energy
up to 20~GeV.  

This test beam setup in the ESA has been used
for the GLAST satellite mission, whose objective is to study
energetic astrophysical gamma rays with energies in the 20~MeV to TeV range. 
The GLAST detector package consists of sophisticated silicon tracker and CsI
calorimeter. It is important to calibrate its response and understand 
the various analysis algorithms in a controlled test beam
environment before its space launch.  Details on this experiment
can be found elsewhere\cite{GLAST}.

\subsection{X-ray Spectroscopy}
Recent X-ray observations of AGN galaxies have revealed features in the iron
emission lines that are characteristic of Doppler shifts and gravitational
redshifts expected from accretion disk models.\cite{tanaka}
The emission lines can be thought of as ``clocks'' moving in various
circular orbits around the black hole.
To further probe the spacetime structure in the accretion disk, high
resolution imaging and broad-band spectroscopy, such as those planned for
the Constellation-X and MAXIM missions, are needed.
A detailed laboratory measurement of heavy element atomic transitions 
and associated polarization effects will also be required for a proper 
interpretation of the observational data.\cite{begelman}

For this purpose,
an intense X-ray source, such as those available at a 
synchrotron light-source facility would be valuable.  The next generation
of linac-based light-source, with
peak brilliance in the range of $10^{32}$-$10^{33}$
photons/sec/mrad$^2$/mm$^2$/0.1\% bandwidth at 1-10 keV, could also play a
role in this study.

\subsection{Air Fluorescence Efficiency Measurement}

The study of ultrahigh-energy cosmic rays has been 
based on observations of the
secondary shower particles resulting from interactions in
the atmosphere.  
For cosmic ray energies above $\sim$10$^{14}$~eV, the shower particles
can reach ground level and extend over a large area.  One observation technique
uses an array of sparsely spaced ground
detectors to measure the density of these shower particles, which
is related to the energy of the primary cosmic ray.  The Akeno
Giant Air Shower Array (AGASA) near Tokyo, Japan, for example,
covers an area of approximately 100~km$^2$, with 100 detector units
separated by about 1~km from each other.\cite{akeno}

The cosmic ray shower also generates a trail of fluorescent light. 
The fluorescence is emitted nearly isotropically, mostly by the
nitrogen molecules in air excited by shower secondaries.
Instead of studying the transverse profile of the shower, as
in the ground array approach, fluorescence-based detectors
use a system of mirrors and photomultipliers to image the shower's longitudinal
development.  The fluorescence luminosity is related to
the primary's energy; and the shape of the longitudinal profile provides
information on the primary's composition.  This technique is used by
the Fly's Eye detector and its upgraded version, the 
High Resolution Fly's Eye (HiRes).\cite{flyseye}

Studies of UHECR events showed that
they are not related to any known galactic sources.  If they originated
in extra-galactic sources, interactions with the cosmic microwave 
background radiation would result in the attenuation of their energy.  The
flux above 10$^{19}$~eV is expected to drop significantly due to
the production of pions -- the so-called GZK cutoff.\cite{gzk}  
However, the Fly's Eye/HiRes and AGASA experiments have observed
events greater than 10$^{20}$ eV, well above the GZK cutoff.  
The two experiments have now accumulated similar
exposure at the highest energies.  With increased statistics,
differences between the two measurements have become apparent.  In
particular, the flux measured by HiRes is systematically lower than
that reported by AGASA above $4\times10^{18}$ eV; there is also a 
difference in the energy at which
the observed power-law spectrum changes slope, the so-called ``ankle''
structure.\cite{jnm}  This can be due to tails in the energy resolution
function or other systematic errors, and is currently being investigated 
by both experiments.  

One possible contribution to this discrepancy is the air fluorescence yield.
Current understanding of air fluorescence, based on previous measurements,
is incomplete.
Many issues still remain: the detailed shape of the fluorescence
spectrum, the pressure and atmospheric impurities dependences, and the
dependence of fluorescence yield on shower particle energy.
The associated systematic uncertainty is estimated to be 15\%. 
A more precise measurement
is desired as improvements are being made to other systematics in
the observation.

\begin{figure}[tb]
\centerline{\epsfxsize=3.3in\epsfbox{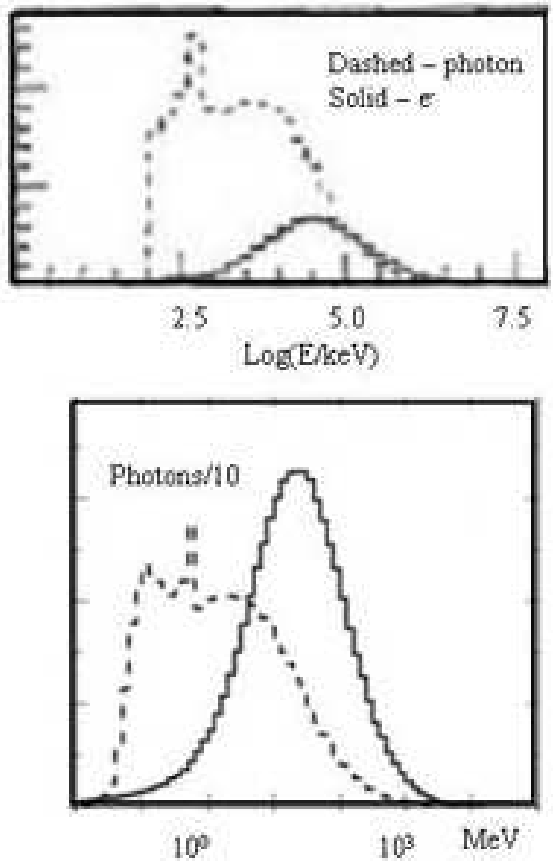}}   
\caption{Simulations of shower secondary electron and photon energy 
distributions for
a $10^{17}$ eV cosmic ray proton (top, CORSIKA) 
and a 28.5 GeV electron beam (bottom, GEANT)
at the shower maximum.\label{corsika-geant}}
\end{figure}

Recently, it has been suggested 
that the high intensity electron beams at SLAC can be used
for such study.\cite{ps}  At the relevant energies, air
showers produced by a cosmic ray hadron is a superposition of electromagnetic
sub-showers.  At the shower's maximum, it consists of
mostly electrons with energies dissipated to the 100 MeV level, near the
critical energy of air.  Further shower development is dominated by
energy loss through ionization and excitation rather than shower particle
creation.  SLAC's electron beams interacting in an air-equivalent alumina 
target produce similar secondary electron energy distributions
-- see Figure~\ref{corsika-geant}.  
The SLAC E-165 experiment -- Fluorescence in Air from Showers (FLASH) --
has been proposed to study in detail the fluorescence yield in an air shower.
It aims to make precision measurements of the total air fluorescence yield,
as well as the spectral, pressure, composition, and
energy dependencies.  Details on this experiment have been
presented elsewhere at this Workshop.\cite{jnm}  

Other examples of accelerator-based
experiments that support astrophysical investigations are
measurement of the Landau-Pomeranchuk-Migdal (LPM) effect,
which has implication for photon/hadron identification at high energies,
and observation of the Askarian effect, which can be used
to detect UHE neutrinos.  These experiments have been carried out 
using SLAC beams.\cite{lpm-aska}

\section{Relativistic Plasma Experiments}
\label{sec-2}

While important issues remain to be resolved in the observational results 
of super-GZK events, the
existence of extra-galactic UHECR above $10^{18}$ eV is well established.
The nature of these cosmic rays and their acceleration mechanism are
still a mystery, and various models have been proposed as 
solutions.\cite{sciencemag-rev}  
In the so-called ``top-down'' approach, the decay products of massive particles
produced in the early universe could account for the observed UHECRs, 
especially
those above the GZK cutoff.  Certain ``grand-unification'' theories
predict the existence of particles with mass around $10^{16}$ GeV.
Particles more massive than this, if they were to explain 
super-GZK events, would have to be produced continuously 
since their lifetimes would be extremely short.
In some theories these particles can be emitted from topological defects 
created between causally
disconnected regions during early epochs of cosmological phase transitions.

In the ``bottom-up'' approach, conventional particles accelerated in powerful
astrophysical systems are thought to be responsible for the observed UHECR
spectrum.  The acceleration mechanisms are complex, involving
strong magnetic fields and nonlinear plasma effects.  Diffusive shock
acceleration
has been the generally accepted model.\cite{shock-acc}
More recent ideas include unipolar induction acceleration\cite{zevatron} and 
high gradient plasma acceleration in wakefields created by
Alfv\'en shocks\cite{chen-tajima}.
Possible acceleration sites are AGNs and gamma-ray bursts where typically
relativistic plasma outflows are present.  
The key observational feature of UHECR is the power-law spectrum.
The appropriate spectral index is predicted by existing models.
Our goal is to experimentally test some of these models in the laboratory.

Typical beams delivered for experimentation in the FFTB are
in short pulses pico-seconds long, $10~\mu$m in radius,
and consists of $10^{10}$ particles.  Thus, the pulse power is approximately
40~Petawatts, and the intensity is $\sim$10$^{20}$~W/cm$^2$.  
The energy density in the bunch is on the order of $10^{13}$~J/m$^3$.
For comparison, 
the threshold for high-energy-density conditions, the energy density 
in a hydrogen molecule or the bulk moduli of solid-state materials,
is $10^{11}$~J/m$^3$
The strong nonlinear and collective responses of a bunched relativistic
particle beam to external stimuli are some of the important characteristics
of a high-energy-density system relevant for astrophysical studies.
Here we discuss possible relativistic plasma experiments.
In particular, we explore
the possibility of merging electron and positron beams to form a
kinetically relativistic plasma, 
allowing the laboratory investigation of 
cosmic high-energy acceleration and radiation production phenomena.

\subsection{e$^{+}$e$^{-}$ Beams as Relativistic Plasma}
\label{rel-plasma}

Neutral co-moving e$^{+}$e$^{-}$ beams have been investigated in an effort
to improve the luminosity limit in high energy e$^{+}$e$^{-}$ storage ring
colliders.
The disruptive effect of one beam's electromagnetic fields on the other
can be compensated, in principle, by colliding neutral beams.  This idea
had been studied using two pairs of 0.8~GeV beams.\cite{leDuff}
The experiment demonstrated 
beam-charge compensation with improved luminosity.

\begin{figure}[tb]
\centerline{\epsfxsize=4.5in\epsfbox{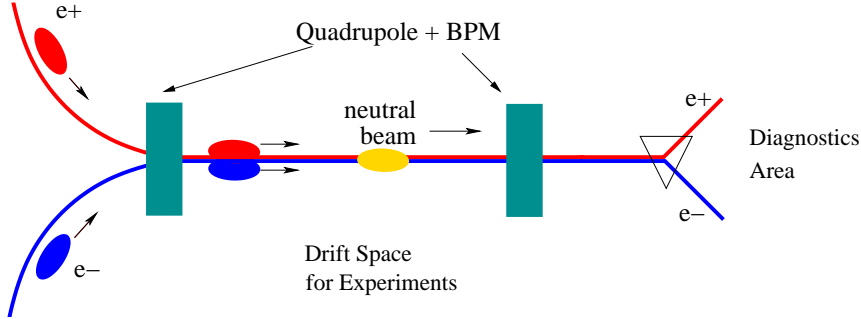}}   
\caption{Combining electron and positron bunches to form a 
relativistic plasma. \label{asset}}
\end{figure}

For our purpose, the 1-GeV 
electron and positron beams emerging from the damping
rings at the beginning of the SLAC linac (see Figure~\ref{slac}) could be
combined, forming an e$^{+}$e$^{-}$ 
plasma streaming at relativistic velocity.\cite{ng}
The transverse positions of the two beams would be aligned to 
micron precision using high resolution beam position monitors.  The temporal
locations would be synchronized using the damping rings' RF phase control,
which is stable at the sub-picosecond level. This level of precision 
beam control has been demonstrated
in measurements of wake-fields in accelerator structures.\cite{asset}
The concept is illustrated in Figure~\ref{asset}.

For a relativistic bunched beam, temperature can be defined in terms of
its emittance.  Analogous to entropy, emittance is a measure of disorder.  
The discussions here follow those in Lawson\cite{lawson}.
The beam's transverse temperature is given by 

\begin{equation}
   kT_{\perp} = \frac{\beta^2\gamma m c^2 \epsilon^2}{4\sigma_r^2}
\end{equation}

\noindent
where $\epsilon$ is the beam's emittance, $\gamma$ the Lorentz factor,
and $\sigma_r$ the transverse beam size.  The longitudinal temperature due
to energy spread is negligible for relativistic beams.

The other plasma parameters can now be calculated.
Results are shown in Table~\ref{params}, using typical SLAC parameters at
the exit of the damping rings.  Plasma parameters are given in
the frame co-moving with the beams.  As can be seen, the number of particles
inside a ``Debye sphere'' ($N_D$) is much greater than one, 
so that the effects of 
individual particles on each other are negligible compared to the 
collective effects, and the plasma description is indeed appropriate.  

For typical AGN jet parameters, the plasma length scales are much
smaller than the jet dimensions.  Thus, the AGN jet plasma is usually
treated as having infinite extend.  For typical relativistic bunched beams, 
however, the Debye radius ($\lambda_D$)
is smaller than the bunch length but larger than the transverse beam
size.  As a consequence, the perpendicular plasma waves 
(involving particle motion in
the transverse direction) have different properties compared to
those excited in an infinite plasma.  However, properties of 
the parallel propagating waves remain the same as those in an infinite
plasma.  The laboratory e$^+$e$^-$ plasma
discussed here can thus model the parallel propagating waves in an infinite
plasma.  As discussed below, this mode is most relevant for
AGN jet dynamics.

\begin{table}[t]
\begin{center}
\caption{Beam parameters in  the laboratory frame 
and corresponding plasma parameters in the co-moving
frame.}
\vspace {1pt}
%
{\footnotesize
\begin{tabular}{|r|l||r|l|}
\hline
Beam       &             & Plasma     &  \\
Parameters &              & Parameters &  \\
(Lab frame) & Value       & (Co-moving frame) & Value \\
\hline

%

Energy (E)    & 1.19 GeV & 
                             Density    &  $4\times10^{11}$ cm$^{-3}$  \\
 $\sigma_E/E$ & $10^{-3}$ & 
                             Debye Length $(\lambda_D)$ & 1.7 mm \\
Bunch Length   & 600 $\mu$m & 
                             Plasma Parameter ($N_D$) & $6 \times 10^6$ \\
Bunch Radius    & 50 $\mu$m &
                    Frequency ($\omega_p/2\pi$)  & $6 \times 10^{9}$ (Hz) \\
Intensity      & $2\times 10^{10}$ & 
                             Wavelength ($\lambda_p$) & 50 mm \\
Density  & $10^{15}$ cm$^{-3}$ & 
                             Skin depth $(c/\omega_p)$ & 8 mm \\
Emittance: & & Temperature: & \\
$\epsilon_x$ & $1.3\times 10^{-8}$ m-rad & 
                             Transverse $(kT_{\perp})$ & 23 keV\\
$\epsilon_y$ & $6.4\times 10^{-10}$ m-rad & 
                              Longitudinal & 0.3 eV \\

\hline
\end{tabular}\label{params}}
\end{center}
\end{table}

So far our discussion have concentrated on neutral plasmas.
The composition of astrophysical jets is, however, far from being understood.
Magnetic confinement is generally accepted as the collimation mechanism,
but it is also highly unstable.
Models of current-carrying jets provide a possible alternative
mechanism where the self-magnetic fields create a pinching force.
This is very similar to the plasma-lens effect familiar to the
beam-plasma physics community.
Non-neutral plasma instabilities relevant for AGN jets 
could be studied using charged
beams readily available at a facility such as SLAC.
Possible experiments are under study.

\subsection{Scaling Laws and Relevance to Astrophysics}

The challenge for laboratory astrophysics
is to create a terrestrial setting which can be scaled 
to the astrophysical environment.  
Magnetohydrodynamic (MHD) models have been used to describe many
astrophysical processes
such as bow-shock excitation in AGN jets or supernova explosions.
The MHD equations have the property that they are 
invariant under the appropriate scale transformations.  This has been the
basis, for example, for designing laser experiments to simulate supernova 
remnants.\cite{remington-takabe,Ryukov,drake}  

The MHD models are applicable when certain assumptions are satisfied.
These, however, may not be applicable to the astrophysical conditions of
interest here.
In the following, we discuss a more general approach based
on kinetic plasma theory.  In particular, we concentrate on
astrophysical plasma processes that might
be investigated using high-energy-density particle beams.  

The observed non-thermal radiation spectrum from AGNs is the subject of
many recent studies.  In some models, broad-band Blazar emission has been 
attributed to synchrotron radiation and/or various forms of 
Compton processes.\cite{sikora} While in other models, 
it is described by the production of 
photon-pairs from the decay of mesons produced via the interaction
of energetic protons with ambient photon and/or matter.\cite{mannheim}  
These models successfully describe
various features in the observed spectrum, and thus are useful for
understanding the radiation processes.  But such phenomenological
approach does not describe details of the underlying micro-physical 
dynamics of AGN jets.  In particular, it does not address the issue of how the 
relativistic jet gives rise to energetic electrons and/or protons
which subsequently produce the radiation.
For example, these models typically assume that
diffusive shock acceleration produces the required power-law spectrum.

In the plasma physics approach, details of the underlying dynamics
for transferring kinetic energy in the relativistic jet into
radiation are described.
In the model proposed by Schlickeiser {\it et al.}\cite{schlik},
the jet is described by a one-dimensional outflow 
consisting of electron and positron pairs with
bulk relativistic velocity, directed parallel to a uniform background
magnetic field.  The pairs have non-relativistic temperature in
the co-moving frame.  The e$^+$e$^-$ jet propagates into an interstellar
medium consisting of cold protons and electrons.

This two-stream multi-fluid system is studied in the jet rest frame.
The analysis starts with a general phase space distribution, and
the calculations then give the dispersion relations of the 
parallel propagating electrostatic (longitudinally polarized)
and low-frequency transverse (Alfven-type) plasma waves.
These waves are excited via a two-stream instability in the pair plasma.
For typical AGN parameters, the calculations show that
the jet kinetic energy is transferred via plasma turbulence to the
initially cold interstellar protons and electrons, which then reach 
a plateau distribution in momenta.  The resulting radiation spectrum
is consistent with observation.\cite{pohl00}

These kinetic plasma calculations also show
that the instability build-up times and growth rates {\it scale}
with the densities and the bulk relativistic factor, while the damping
rates scales also with temperature.  For example, the time it takes
to build up the transverse instability in the protons is given by
$t_{t,p} \sim (1/\omega_{p,e})(n_j/n_i)(m_p \Gamma / m_e)^{4/3}$,
where $\omega_{p,e}$ is the electron plasma frequency, $n_j$ and $n_i$
are the jet and interstellar plasma densities, and $\Gamma$ is the bulk
Lorentz factor.  The Landau damping rate is found to scale with
$\Theta^{3/2} \omega_{p,e} \Gamma^2 {\rm exp} [-(\Gamma-1)/\Theta]$, where
$\Theta = kT/m_e c^2$ is the dimensionless temperature parameter.

\subsection{Parameters for Laboratory Experiments}

\begin{figure}[t]
%
\centerline{\epsfxsize=4.5in\epsfbox{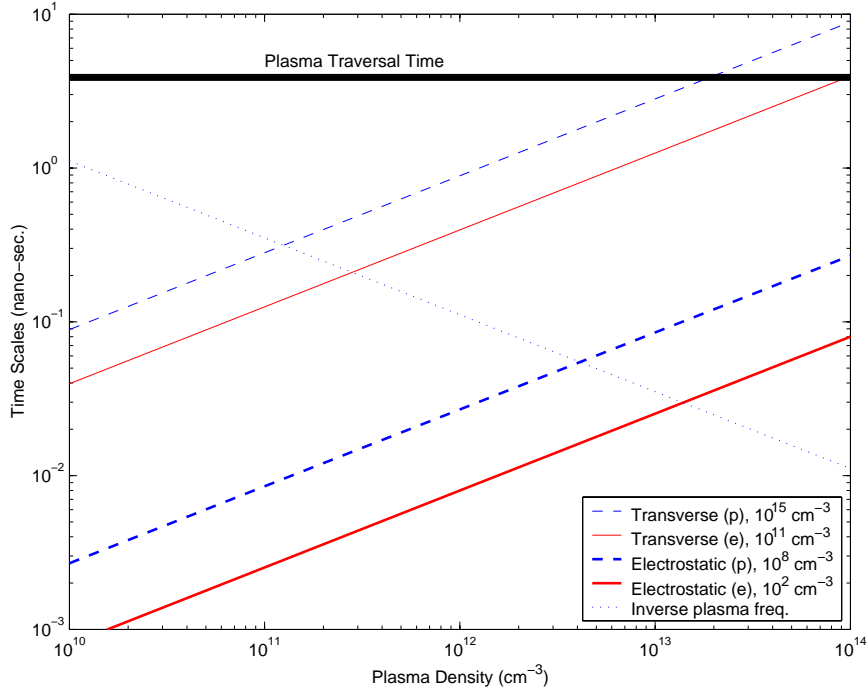}}   
\caption{Various time scales for a laboratory experiment to test the
dynamical model of AGN jets.  The scaling of the plasma wave build-up
times are shown as a function of the jet plasma density, for various
interstellar medium densities -- see text.}
\label{dyn-times}
\end{figure}

To determine whether the parameters of the relativistic plasma
created by merging electron and positron bunches
are relevant for an experimental investigation of AGN dynamics,
the various dynamical time scales are calculated.
The results are shown in Figure~\ref{dyn-times} for the parameters
given in Table~\ref{params}.  In the setup being considered here, 
the pair plasma in the co-moving frame appears to be $\sim$1-m long to the 
ambient plasma traveling through it.
As can be seen from Figure~\ref{dyn-times},
all dynamical time scales are shorter than the plasma traversal time:
the time during which the relativistic plasma and the ambient plasma
interact with each other.  
Typical plasma time scales are shown as the inverse plasma frequency.
The build-up of the electrostatic waves is rather quick, for both the
electrons and the protons, even with a fairly thin ambient plasma.
The build-up of the transverse waves takes much more time, particularly
for the protons, in which case an ambient plasma density of
10$^{15}$~cm$^{-3}$ in the laboratory is required.

Also, the maximum
growth rate of the electrostatic turbulence is much greater than the
Landau damping rate; similarly, the transverse turbulence growth rate is much
larger than the cyclotron damping rate.  Thus,
this set of experimental parameters is in a
regime where strong nonlinear plasma turbulence similar to 
those excited in AGN jets can be
created and studied in detail experimentally in the laboratory.
Further theoretical calculations
are needed to guide the design of the experiment. A
detailed numerical simulation using particle-in-cell
techniques is needed as the next step.  

The transverse magneto-hydrodynamic (Alfven-type) wave is
especially interesting for testing various cosmic acceleration mechanisms.
This type of turbulence is crucial in the formation of collisionless shocks
and for efficient particle deflection
in the diffusive shock acceleration process.
The Alfven wave is also expected to excite plasma wakefields,
which can provide high gradient particle acceleration.  
The spectrum and polarization properties of the radiation produced
in the interaction of this e$^+$e$^-$ ``jet'' with an ambient plasma
can be measured and compared with astrophysical observations.
Detailed simulation studies for these experiments are underway.

The laboratory experiments described here could have applications
beyond the understanding of AGN jet dynamics.
The dynamics in the polar caps of a spinning neutron star
have been studied in the context of
relativistic streaming electron-positron plasma.\cite{lyutikov} 
Also, if GRB radiation is beamed, its dynamics would be similar to those
found in AGN jets.  Thus, our laboratory experiments would also shed
light on these systems.

\section{Summary and outlook}
\label{sec-3}

The field of laboratory
astrophysics holds promise to the understanding of some of the most
exciting astrophysical observations today.  
%
%
We have shown that
particle accelerators are excellent tools for laboratory astrophysics,
providing calibration data for observations and bench-marking
computer models, as well as creating extreme conditions that make possible
investigation of astrophysical dynamics in 
a terrestrial laboratory.  SLAC, with the existing
expertise and infrastructure, is well-positioned to
contribute to this rapidly growing field.\cite{workshops}
The proposed ORION\cite{orion-web} facility
for advanced accelerator research and beam physics
will also be able to support dedicated laboratory astrophysics experiments
with its unique combination of high quality electron beams and
diagnostic lasers. 

\section*{Acknowledgments}
I would like to thank the Workshop organizers for their kind hospitality.
I would also like to thank P. Chen, R. Noble, K. Reil, and 
P. Sokolsky for many fruitful discussions.  





%
%
%
%

\end{document}